\documentclass[aps,prl,twocolumn,superscriptaddress,amsfont,graphicx]{revtex4}%
\usepackage{color,graphicx,epsfig}
\usepackage{ifpdf}
\usepackage{amsmath}
\usepackage{bm}
\usepackage{color}
\usepackage[english]{babel}
\usepackage{graphicx}%
\usepackage{amsfonts}%
\usepackage{amssymb}
\usepackage{braket}
\usepackage{hyperref}
\usepackage{enumerate}

\definecolor{nicered}{rgb}{0.7,0.1,0.1}
\definecolor{nicegreen}{rgb}{0.1,0.5,0.1}
\hypersetup{colorlinks,citecolor= nicegreen,linkcolor= nicered}

\newcommand{\beq}{\begin{equation}}
\newcommand{\eeq}{\end{equation}}
\newcommand{\bea}{\begin{eqnarray}}
\newcommand{\eea}{\end{eqnarray}}

\definecolor{Red}{rgb}{1.,0.,0.}

\def\taun{{\cal T}_N}

\def\tauncut{{\cal T}_N^{cut}}
\def\tauonecut{{\cal T}_1^{cut}}

\def\mysection#1{{{\bf #1}.~}}
\def\OMIT#1{}

\begin{document}

\def\Peking{Center for High-Energy Physics, Peking University, Beijing, 100871, China}
\def\Maryland{Maryland Center for Fundamental Physics, University of Maryland, College Park, Maryland 20742, USA}
\def\Argonne{High Energy Physics Division, Argonne National Laboratory, Argonne, IL 60439, USA}
\def\Northwestern{Department of Physics \& Astronomy, Northwestern University, Evanston, IL 60208, USA}


\title{$W$-boson production in association with a jet at next-to-next-to-leading order in perturbative QCD}

\author{Radja Boughezal}     
\email[Electronic address:]{rboughezal@anl.gov}
\affiliation{\Argonne}

\author{Christfried Focke}
\email[Electronic address:]{christfried.focke@northwestern.edu}
\affiliation{\Northwestern}

\author{Xiaohui Liu}
\email[Electronic address:]{xhliu@umd.edu}
\affiliation{\Maryland}
\affiliation{\Peking}

\author{Frank Petriello}     
\email[Electronic address:]{f-petriello@northwestern.edu}
\affiliation{\Argonne}
\affiliation{\Northwestern}

\date{\today}
\begin{abstract}
We present the complete calculation of $W$-boson production in association with a jet in hadronic collisions through next-to-next-to-leading order in perturbative QCD.  To cancel infrared divergences we discuss a new subtraction method that exploits the fact that the $N$-jettiness event-shape variable fully captures the singularity structure of QCD amplitudes with final-state partons.  This method holds for processes with an arbitrary number of jets, and is easily implemented into existing frameworks for higher-order calculations.  We present initial phenomenological results for $W$+jet production at the LHC.  The NNLO corrections are small and lead to a significantly reduced theoretical error, opening the door to precision measurements in the $W$+jet channel at the LHC.

\end{abstract}

\maketitle

\section{Introduction} \label{sec:intro}

Run I of the Large Hadron Collider (LHC) was a remarkable success, culminating in the 2012 discovery of the long-awaited Higgs boson.  One major contributor to this exciting outcome was the precision QCD framework used to model and understand both the sought-after signals and the often overwhelming backgrounds.  
The demand for ever-more sophisticated precision QCD calculations is only increasing as Run II of the LHC begins.

Calculations through next-to-next-to-leading order (NNLO) in perturbative QCD are becoming increasingly necessary to match the precision of LHC measurements.  
%
During the past years the first few NNLO calculations for $2 \to 2$ scattering processes, needed for precision phenomenology in many channels at the LHC, have begun to appear.  The full results for $t\bar{t}$ production~\cite{Czakon:2013goa}, single-top production~\cite{Brucherseifer:2014ama}
and for several color-neutral final states are available~\cite{Catani:2011qz}.  The situation is less advanced for processes containing final-state jets, which possess a more complex singularity structure.  Partial results are available for inclusive jet production~\cite{Ridder:2013mf} and Higgs+jet production~\cite{Boughezal:2013uia}.  All such NNLO calculations require a subtraction scheme to extract infrared singularities from real-emission amplitudes and cancel them against the corresponding divergences appearing in the virtual amplitudes.  Currently, two subtraction schemes have been demonstrated to successfully handle processes containing final-state jets in hadronic collisions: antennae subtraction~\cite{GehrmannDeRidder:2005cm} and sector-improved residue subtraction~\cite{Czakon:2010td}.  Although powerful, both approaches require the development of a sophisticated and specialized machinery.  It appears difficult to merge such calculations into a common framework with other standard tools used by the high energy physics community.

An alternative to such techniques is $q_T$-subtraction~\cite{Catani:2007vq}.  It leverages the observation that for color-neutral final states, the transverse momentum $q_T$ of the final  state completely determines the singularity structure of the contributing QCD amplitudes.   For $q_T$ above some small resolution parameter, the result is simply the NLO calculation for the color-neutral state plus an additional jet.  Below the cutoff, the cross section is dominated by large logarithms of the final-state mass scale over $q_T$.  This cross section can be obtained by expansion of the well-known CSS resummation formula~\cite{Collins:1984kg} to fixed order.  Unfortunately, $q_T$-subtraction cannot be extended beyond color-neutral final states, since the transverse momentum no longer completely describes the singularity structure of QCD cross sections with final-state jets. 

In this manuscript we discuss a subtraction scheme that overcomes these various limitations.  It is valid for an arbitrary number of final-state jets, and maximally reuses information available in existing NLO calculations.  It is based on the observation that the $N$-jettiness event shape variable $\taun$~\cite{Stewart:2010tn} controls the singularity structure of QCD cross sections with $N$ final-state jets.  One can partition the phase space according to a resolution parameter $\tauncut$.  Above the cutoff, the result is the NLO calculation with $N+1$ jets.  Below the cutoff, the resummation formula for $\taun$ can be used to obtain the cross section through NNLO.  In this phase-space region the cross section can be written in terms of a small number of known universal functions, together with the process-dependent two-loop virtual corrections.  A similar division of the final-state phase space using the invariant mass of the hadronic radiation was used to obtain top-quark decay at NNLO~\cite{Gao:2012ja}; we discuss here the general applicability of this idea to $N$-jet cross sections through the use of $\taun$.

  We demonstrate our method with a highly non-trivial example: $W$+jet at NNLO.  This calculation is needed for numerous LHC phenomenological applications, including the determination of the gluon distribution function. 
We present here selected numerical results for $W$+jet production for the LHC.  
We find that the NNLO corrections decrease the NLO cross section by approximately 1\%, and significantly reduce the residual theoretical uncertainty as estimated by scale variation.  Our computation makes precision measurements possible in the $W$+jet process at the LHC.

\section{Theoretical Framework} \label{sec:theory}

We sketch here the construction of the $N$-jettiness subtraction scheme. We begin with the definition of $N$-jettiness$,\taun$, a global event shape variable designed to veto final-state jets~\cite{Stewart:2010tn}:
\begin{equation}
\label{eq:taudef}
{\cal T}_N = \sum_k \text{min}_i \left\{ \frac{2 p_i \cdot q_k}{Q_i}\right\}.
\end{equation}
The subscript $N$ denotes the number of jets desired in the final state, and is an input to the measurement.  For the $W$+jet process considered here, we have $N=1$.  Values of ${\cal T}_1$ near zero indicate a final state containing a single narrow energy deposition, while larger values denote a final state containing two or more well-separated energy depositions.  
The $p_i$ are light-like vectors for each of the initial beams and final-state jets in the problem, while the $q_k$ denote the four-momenta of any final-state radiation.  The $Q_i$ are dimensionful variables that characterize the hardness of the beam-jets and final-state jets.  We set $Q_i = 2 E_i$, twice the energy of each jet.  

To proceed, we first note that at NNLO, the cross section consists of contributions with Born-level kinematics, and processes with either one or two additional partons radiated.  We partition the phase space for each of these terms into regions above and below $\tauncut$:
\begin{equation}
\label{eq:partition}
\begin{split}
\sigma_{NNLO} &= \int {\rm d}\Phi_N \, |{\cal M}_{N}|^2 +\int {\rm d}\Phi_{N+1} \, |{\cal M}_{N+1}|^2 \, \theta_N^{<} \\
&+\int {\rm d}\Phi_{N+2} \, |{\cal M}_{N+2}|^2 \, \theta_N^{<}+\int {\rm d}\Phi_{N+1} \, |{\cal M}_{N+1}|^2 \, \theta_N^{>} \\
&+\int {\rm d}\Phi_{N+2} \, |{\cal M}_{N+2}|^2 \, \theta_N^{>} \\
& \equiv  \sigma_{NNLO}(\taun < \tauncut)+\sigma_{NNLO}(\taun > \tauncut)
\end{split}
\end{equation}
Here, we have abbreviated $\theta_N^{<} = \theta(\tauncut-\taun)$ and $\theta_N^{>} = \theta(\taun-\tauncut)$, and have suppressed for simplicity the allowed introduction of any infrared-safe measurement function under the phase-space integral.  The first three terms in this expression all have $\taun<\tauncut$, and have been collectively denoted as $\sigma_{NNLO}(\taun < \tauncut)$.  The remaining two terms have $\taun>\tauncut$, and have been collectively denoted as $\sigma_{NNLO}(\taun > \tauncut)$.  Contributions with Born-level kinematics necessarily have $\taun=0$.  We note that a similar partitioning of phase space to separate the singular and non-singular regions in $\taun$ has proven quite useful in the context of merging fixed-order calculations with parton showers in the effective-theory framework pioneered by the \textsc{Geneva} collaboration~\cite{Alioli:2012fc}.

The key advance that allows us to compute the cross section to NNLO below $\tauncut$ is the existence of a factorization theorem that gives an all-orders description of $N$-jettiness for small $\taun$~\cite{Stewart:2009yx,Stewart:2010pd}.  
Using this factorization theorem, the cross section for a hadronic process with $\taun$ less than some value $\tauncut$ can be written in the schematic form
\begin{equation} \label{eq:fact}
 \sigma(\taun < \tauncut)=  \int H \otimes B \otimes B \otimes S \otimes   \left[ \prod_{n}^{N} J_n \right] +\cdots .
\end{equation}
Here, $H$ describes the effect of hard radiation; when dimensional regularization is used this function simply encodes the virtual corrections to the process.  $B$ encodes the effect of radiation collinear to one of the two initial beam directions; the importance of the beam function more generally in describing hadronic collisions was first pointed out in Ref.~\cite{Stewart:2009yx}.  It can be further decomposed as a perturbative matching coefficient convolved with the usual parton distribution function.  $S$ describes the soft radiation, and $J_n$ contains the radiation collinear to a final-state jet.  
The ellipsis denotes power-suppressed terms which become negligible for $\taun \ll Q_i$. The derivation of this all-orders expression in the small-$\taun$ limit relies heavily upon the machinery of Soft-Collinear Effective Theory~\cite{Bauer:2000ew}.

If this formula is expanded to fixed-order in the strong coupling constant, it reproduces the fixed-order cross section $\sigma_{NNLO}(\taun < \tauncut)$ for low $\tauncut$ needed in Eq.~(\ref{eq:partition}).  Currently the hard function is known at NNLO for many phenomenologically interesting cases, including the $W$+jet process considered here~\cite{Gehrmann:2011ab}.  The beam functions are known at NNLO~\cite{Gaunt:2014xga}, as are the jet functions~\cite{Becher:2006qw} and the soft function~\cite{Boughezal:2015eha}. 

A full NNLO calculation requires as well the high $\taun$ region above $\tauncut$.  However, a finite value of $\taun$ implies that there are actually $N+1$ resolved partons in the final state.  This is the crucial observation; $\taun$ completely describes the singularity structure of QCD amplitudes that contain $N$ final-state partons at leading order.  The high $\taun$ region of phase space is therefore described by a NLO calculation with $N+1$ jets.  
We must choose $\tauncut$ much smaller than any other kinematical invariant in the problem in order to avoid power corrections to Eq.~(\ref{eq:fact}) below the cutoff.  
We address this issue explicitly in the context of $W$+jet in the next section.  

To summarize, we list here the exact steps needed for our calculation of the full NNLO calculation of $W$+jet.
\begin{itemize}

\item Generate an event for the $N+1$-jet process at NLO, which may contain either $N+1$ partons or $N+2$ partons.  Determine the reference vectors $p_i$ in Eq.~(\ref{eq:taudef}) by performing a pre-clustering of the radiation using a jet algorithm, as discussed in Refs.~\cite{Stewart:2010tn,Jouttenus:2011wh}.  The determination of the $p_i$ is insensitive to the choice of jet algorithm in the small-$\tauncut$ limit~\cite{Stewart:2010tn}.

\item Calculate $\taun$ according to Eq.~(\ref{eq:taudef}).  If $\taun>\tauncut$, keep the event.  Events satisfying this criterion form the NLO cross section for the $N+1$-jet process $\sigma_{NNLO}(\taun > \tauncut)$ needed in Eq.~(\ref{eq:partition}).  This NLO cross section can be obtained using any standard technique or code. If $\taun<\tauncut$, reject the event.

\item Obtain the cross section $ \sigma(\taun < \tauncut)$ by expanding the resummation formulae of Eq.~(\ref{eq:fact}) to NNLO.  We note that this fully accounts for all $N$-parton contributions in Eq.~(\ref{eq:partition}), as well as for terms with additional partons where $\taun < \tauncut$.

\end{itemize}

\section{Validation of the Formalism} \label{sec:formalism}

We next discuss the validation of the $\taun$-subtraction formalism, in the context of our calculation of $W$+jet at NNLO.  An advantage of our formalism is that it maximally reuses known information coming from existing NLO calculations.  Above $\tauncut$ we need a NLO calculation of $W$+2-jets, which we obtain from MCFMv8.0~\cite{Boughezal:2016wmq}.  For the terms which contribute below $\tauncut$, we have checked our implementation of the two-loop virtual corrections against those contained in PeTeR~\cite{Becher:2011fc}.  Our calculation and validation of the necessary $N$-jettiness soft function has been detailed in a separate publication~\cite{Boughezal:2015eha}.

\begin{figure}[h]
\centering
\includegraphics[width=3.3in]{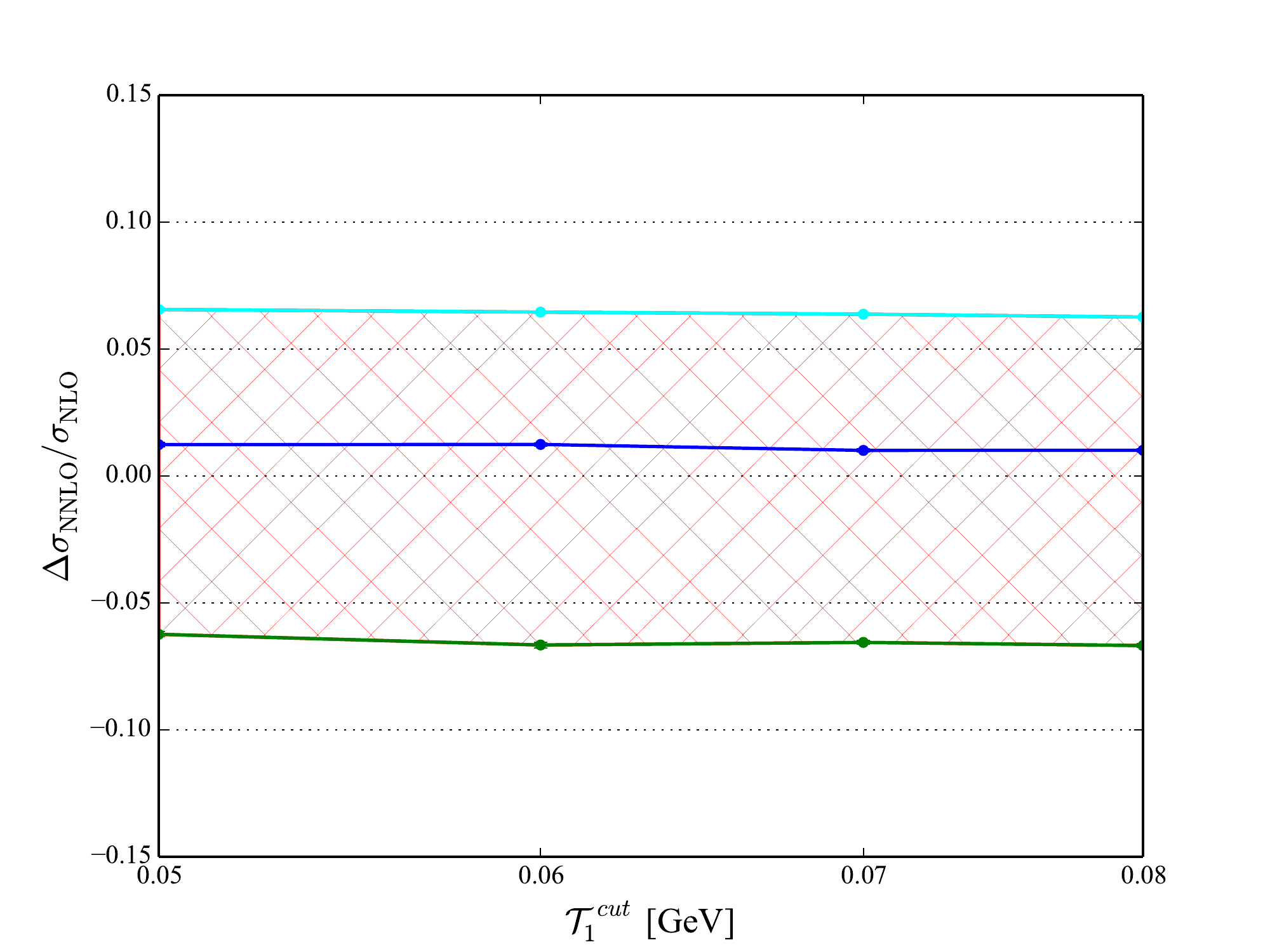}
\caption{The NNLO correction $\Delta \sigma_{\text{NNLO}}$ normalized to the NLO cross section as a function of $\tauonecut$.  The solid blue line denotes the result for the central scale choice $\mu=M_W$, while the red hatched area indicates the correction in the range $M_W/2 \leq \mu \leq 2 M_W$.  The solid cyan and green lines indicate the results for $\mu=2 M_W$ and $\mu=M_W/2$, respectively.  The NNLO result for each scale choice has been normalized to the NLO cross section obtained using the same scale choice.} \label{fig:taucheck}
\end{figure}

A powerful check of our formalism is that the logarithmic dependence on $\tauncut$ that occurs in the separate low and high $\taun$ regions cancels when they are summed.  
We show in Fig.~\ref{fig:taucheck}
the results of this check.  The plot shows only the ${\cal O}(\alpha_s^3)$ correction to the cross section as a function of $\tauncut$; we have checked that the ${\cal O}(\alpha_s^2)$ NLO cross section obtained with this technique agrees exactly with the known results.  These cross sections are obtained using CT14 parton distribution functions~\cite{Dulat:2015mca}, and contain the following fiducial cuts on the final-state jet from CMS~\cite{Khachatryan:2014uva}: $p_T^{jet} > 30$ GeV, $|\eta_{jet}|<2.4$.  The ATLAS analysis is similar but with slightly different cuts~\cite{Aad:2014qxa}.  Both the renormalization and factorization scales have been set to $\mu=M_W$ and varied from this choice by a factor of two. Over the region $0.05 \, \text{GeV} < \tauncut < 0.08\, \text{GeV}$, the sum is stable to better than one picobarn, a size which represents an 0.1\% correction to the total cross section.  
The NLO corrections to $W$+2-jets, have been obtained using the {\tt double precision} version of MCFM, both the single-core version and the new multi-core implementation~\cite{Campbell:2015qma}.  We have checked that for larger values of $\tauncut$, the power corrections in Eq.~(\ref{eq:fact}) begin to become important.  
For the numerical results presented in the remainder of this paper we use the choices $\tauonecut=0.05$, 0.06 GeV to cross-check their $\tauncut$ independence.
 
As a final check of our computation, we have applied our formalism to calculate the NNLO corrections to Higgs production in association with a jet, for which partial results are available~\cite{Boughezal:2013uia}.  We find agreement with the fiducial cross section for this process.  The details of this computation, together with phenomenological results for the LHC, will be presented in a separate manuscript~\cite{newhiggs}.  

\section{Numerical Results}

We present here first numerical results for $W^{+}$+jet production at the LHC.  We focus on $\sqrt{s}=8$ TeV collisions, and use CT14 parton distribution functions~\cite{Dulat:2015mca} at the same order of perturbation theory as the corresponding partonic cross section.  Our central scale choice is $\mu=M_W$.  To obtain an estimate of the theoretical errors we vary $\mu$ away from this choice by a factor of two.  We again impose the following cuts on the final-state jet: $p_T^{jet} > 30$ GeV, $|\eta_{jet}|<2.4$.  We reconstruct jets using the anti-$k_T$ algorithm~\cite{Cacciari:2008gp} with $R=0.5$.

We begin by showing in Table~\ref{tab:fiducial} the total cross section subject to the above cuts at LO, NLO, and NNLO in the strong coupling constant.  These numbers include the branching fraction for the $W$-boson to decay to a single lepton flavor.  We note that the numerical error on these numbers is at the several-per-mille level.  The cross section shifts by $+40\%$ when going from LO to NLO in perturbation theory, but only by approximately +1\% when going from NLO to NNLO.  The scale variation is approximately $\pm 7\%$ at LO and NLO, while at NNLO it is reduced to the percent level.  We note that at NNLO the largest cross section is obtained for $\mu=M_W$, leading to the lack of scale variation in the upper direction in the Table below. 
The residual theoretical error is reduced to the percent level at NNLO, and excellent convergence of the perturbative series is obtained.
\begin{table}[h]
\begin{tabular}{|l|l|}
\hline
\multicolumn{2}{|c|}{$p_T^{jet}> 30$ GeV, $|\eta_{jet}|<2.4$}\\ \hline\hline
Leading order: & $561^{+47}_{-35}$ pb \\ \hline
Next-to-leading order: & $796^{+62}_{-49}$ pb \\ \hline
Next-to-next-to-leading order: & $807^{+0}_{-10}$ pb \\ \hline
\end{tabular} \caption{Fiducial cross sections, defined by $p_T^{jet}> 30$ GeV, $|\eta_{jet}|<2.4$, using CT14 PDFs at each order of perturbation theory.  } \label{tab:fiducial}
\end{table}

\begin{figure}[h]
\centering
\includegraphics[width=3.4in]{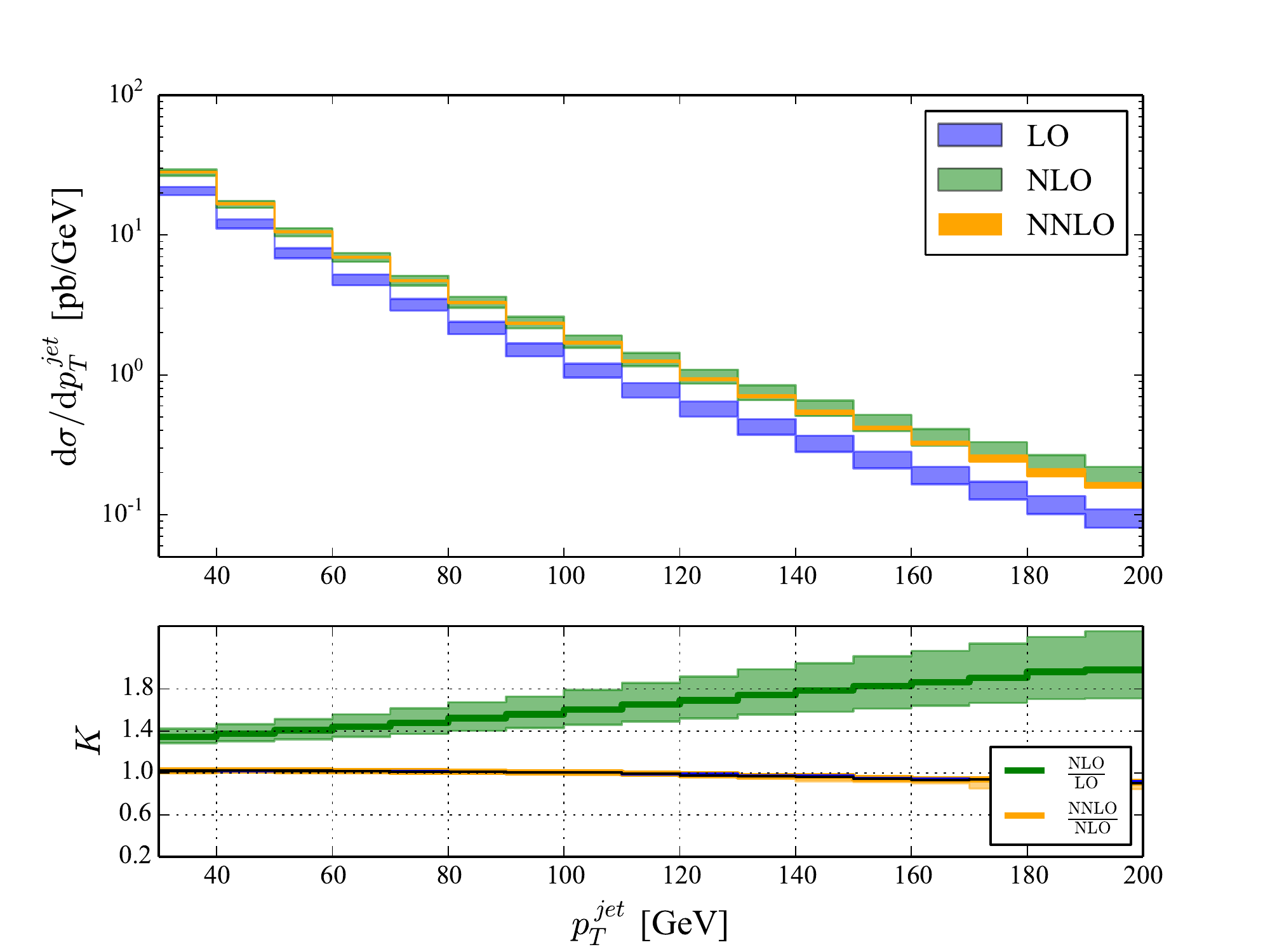}
\caption{The transverse momentum spectrum of the leading jet at LO, NLO and NNLO in perturbation theory.  The bands indicate the estimated theoretical error. The lower inset shows the ratios of the NLO over the LO cross section, and the NNLO over the NLO cross section.  The red vertical error bars in the lower inset indicate the scale-variation error.  The blue and black lines in the lower inset respectively show the distribution for $\tauncut=0.05$ GeV and $\tauncut=0.06$ GeV, for the scale choice $\mu=M_W$.} \label{fig:ptspectrum}
\end{figure}

In Fig.~\ref{fig:ptspectrum} we show the transverse momentum spectrum of the leading jet at LO, NLO and NNLO in perturbation theory.  The ratios of the NLO cross section over the LO result, as well as the NNLO cross section over the NLO one, are shown in the lower inset.  The shaded bands in the upper inset indicate the theoretical errors at each order estimated by varying the renormalization and factorization scales by a factor of two around their central value, as do the vertical error bars in the lower inset.  In the lower inset we have shown the results for both $\tauncut=0.05$ GeV and $\tauncut=0.06$ GeV, for the scale choice $\mu=M_W$, to demonstrate the $\tauncut$ independence in every bin studied.  The NLO corrections are large and positive for this scale choice, increasing the cross section by 40\% at $p_T^{jet}=40$ GeV and by nearly a factor of two at $p_T^{jet}=180$ GeV.  The scale variation at NLO reaches approximately $\pm 20\%$ for $p_T^{jet}=180$ GeV.  The shift when going from NLO to NNLO is much more mild, giving only a percent-level decrease of the cross section that varies only slightly as $p_T^{jet}$ is increased.  The scale variation at NNLO is at the percent level and is nearly invisible on this plot.  

\begin{figure}[h]
\centering
\includegraphics[width=3.4in]{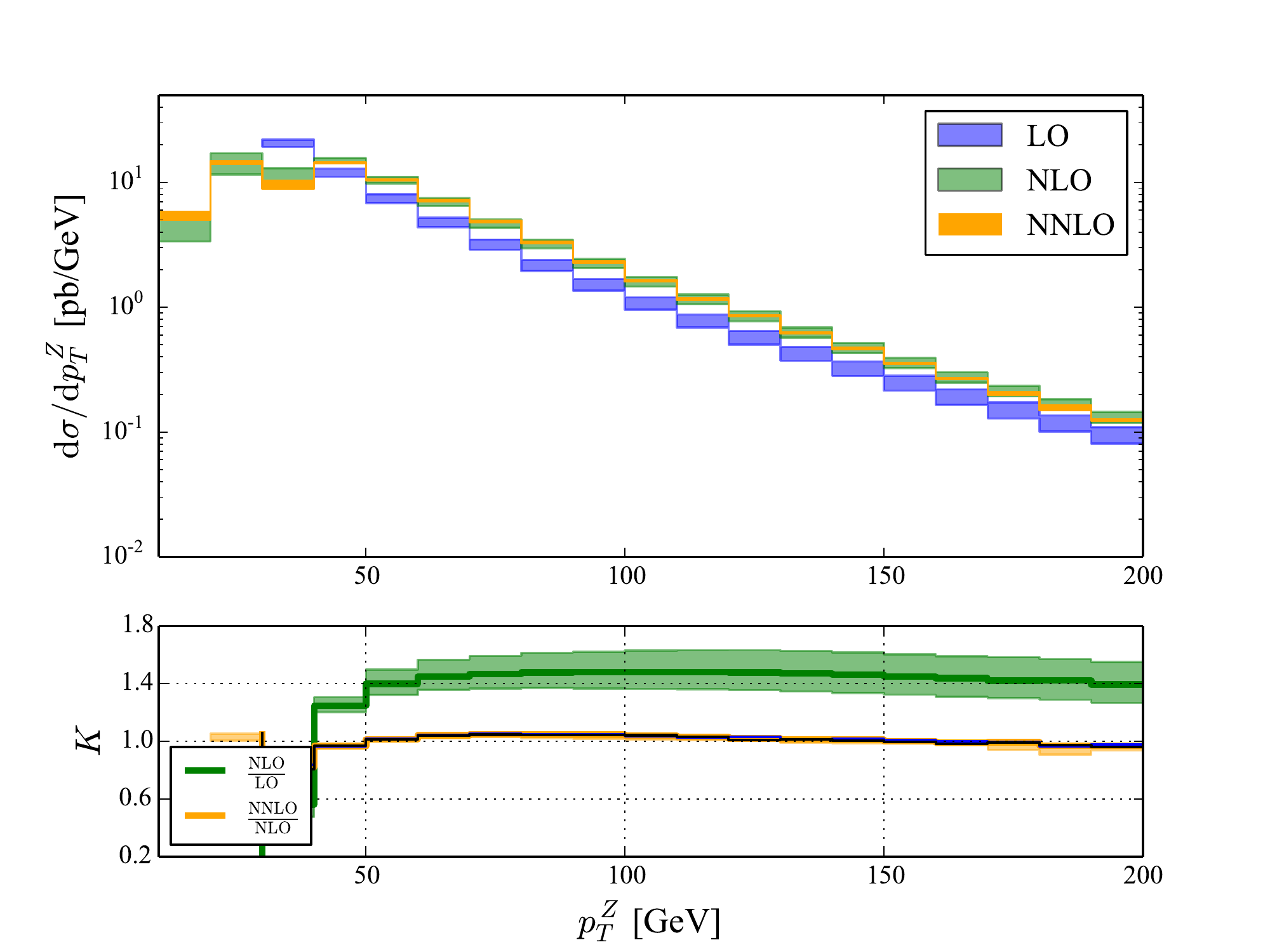}
\caption{The transverse momentum spectrum of the $W$-boson at LO, NLO and NNLO in perturbation theory.  The bands indicate the estimated theoretical error. The lower inset shows the ratios of the NLO over the LO cross section, and the NNLO over the NLO cross section. The red vertical error bars in the lower inset indicate the scale-variation error.  The blue and black lines in the lower inset respectively show the distribution for $\tauncut=0.05$ GeV and $\tauncut=0.06$ GeV, for the scale choice $\mu=M_W$.} \label{fig:ptWspectrum}
\end{figure}

The transverse momentum spectrum of the $W$-boson is shown in Fig.~\ref{fig:ptWspectrum}.  The NLO corrections are again 40\% for $p_T^W \geq 50$ GeV with a sizable scale dependence, while the NNLO corrections are flat in this region and decrease the cross section by a small amount.  The phase-space region $p_T^W <30$ GeV only opens up at NLO, leading to a different pattern of corrections for these transverse momentum values.  The instability of the perturbative series in the bins closest to the boundary $p_T^W=30$ GeV is caused by the well-known Sudakov-shoulder effect~\cite{Catani:1997xc}.

\section{Conclusions}

We have presented in this manuscript the complete NNLO calculation of $W$-boson production in association with a jet in hadronic collisions.  To perform this computation we have discussed a new subtraction scheme based on the $N$-jettiness event-shape variable  $\taun$.  
 We have validated our approach in several ways: when possible the various components have been cross-checked against known results in the literature, the necessary cancellation of the logarithmic $\tauncut$ between the phase-space regions $\taun>\tauncut$ and $\taun<\tauncut$ has been established, and we have reproduced known results for Higgs production in association with a jet at NNLO.  
The NNLO corrections to the $W$+jet process indicate a remarkably stable perturbative series ready to be used for precision measurements at the LHC.  
We will further study the phenomenological impact of our NNLO result in future work, including the prediction for the exclusive one-jet bin, where an intricate interplay between various sources of higher-order corrections was recently pointed out~\cite{Boughezal:2015oga}.

We believe that the development of the jettiness-subtraction technique represents a watershed moment in the field of higher-order calculations.  For the first time a completely general subtraction scheme valid for any number of final-state jets has been introduced that is based on a factorization theorem which extends to all orders in perturbation theory and is straightforward to implement in existing frameworks for NLO calculations.  We anticipate that the $W$+jet process presented here is only the first of many results obtained with this novel technique.
 
\mysection{Acknowledgements}

R.~B. is supported by the DOE contract DE-AC02-06CH11357.  C.~F. is supported by the DOE grant DE-FG02-91ER40684.  X.~L. is supported by the U.S. DOE.  F.~P. is supported by the DOE grants DE-FG02-91ER40684 and DE-AC02-06CH11357.  This research used resources of the National Energy Research Scientific Computing Center, a DOE Office of Science User Facility supported by the Office of Science of the U.S. Department of Energy under Contract No. DE-AC02-05CH11231.

\end{document}